\documentclass[12pt]{article}

\usepackage{times}
\usepackage{amsmath}
\usepackage{amsfonts}
\usepackage{amssymb}
\usepackage{graphicx}
\usepackage{mathrsfs}
\usepackage{braket}
\usepackage{color}

\title{Magneto-transport evidence for strong topological insulator phase in ZrTe$_5$} 

\author
{Jingyue Wang,${}^{1,2}$ Yuxuan Jiang,${}^{3,4\ast}$ Tianhao Zhao,${}^{1}$ Zhiling Dun,${}^{1}$\\ Anna L. Miettinen,${}^{1}$ Xiaosong Wu,${}^{2}$ Martin Mourigal,${}^{1}$ Haidong Zhou,${}^{5}$\\ Wei Pan,${}^{6}$ Dmitry Smirnov,${}^{4}$ Zhigang Jiang${}^{1\ast}$\\
\\
\normalsize{${}^{1}$School of Physics, Georgia Institute of Technology, Atlanta, Georgia 30332, USA} \\
\normalsize{${}^{2}$State Key Laboratory for Artificial Microstructure and Mesoscopic Physics,}\\
\normalsize{Peking University, Beijing 100871, China}\\
\normalsize{${}^{3}$School of Physics and Optoelectronics Engineering, Anhui University,}\\
\normalsize{Hefei, Anhui 230601, China}\\
\normalsize{${}^{4}$National High Magnetic Field Laboratory, Tallahassee, Florida 32310, USA}\\
\normalsize{${}^{5}$Department of Physics and Astronomy, University of Tennessee,}\\
\normalsize{Knoxville, Tennessee 37996, USA}\\
\normalsize{${}^{6}$Quantum and Electronic Materials Department, Sandia National Laboratories,}\\ 
\normalsize{Livermore, California 94551, USA}\\
\\
\normalsize{$^\ast$To whom correspondence should be addressed}\\
\normalsize{ E-mail: yuxuan.jiang@ahu.edu.cn; zhigang.jiang@physics.gatech.edu}
}

\date{}
\begin{document} 


\baselineskip24pt


\maketitle
\clearpage
\section*{Abstract}
The identification of a non-trivial band topology usually relies on directly probing the protected surface/edge states. But, it is difficult to achieve electronically in narrow-gap topological materials due to the small (meV) energy scales. Here, we demonstrate that band inversion, a crucial ingredient of the non-trivial band topology, can serve as an alternative, experimentally accessible indicator. We show that an inverted band can lead to a four-fold splitting of the non-zero Landau levels, contrasting the two-fold splitting (spin splitting only) in the normal band. We confirm our predictions in magneto-transport experiments on a narrow-gap strong topological insulator, zirconium pentatelluride (ZrTe$_5$), with the observation of additional splittings in the quantum oscillations and also an anomalous peak in the extreme quantum limit. Our work establishes an effective strategy for identifying the band inversion as well as the associated topological phases for future topological materials research.\\

\section*{Introduction}

A prime task of topological materials research is to determine the band topology. For topological insulators (TIs), the non-trivial band topology features an inverted gap in the bulk and Dirac-like surface/edge states. The latter has been the smoking gun evidence for identifying TIs in previous studies \cite{TI_review_0,TI_review}. However, finding surface/edge states is challenging near the semiconductor to semimetal transition, where a normal insulator (NI) closes its bandgap and then reopens to form an inverted gap in the TI phase \cite{Murakami_1,Murakami_2}. In this regime, due to the small bandgap (on the meV energy scale), the electronic response of the surface/edge states is buried in that of the bulk. Therefore, accurate determination of the bulk band alignment (normal versus inverted) becomes crucial in decoding the band topology information.

Electronic transport measurement has been a work-horse in exploring topological materials. When combined with a magnetic field ($B$), it reveals important properties of the band structure such as the Fermi surface and Berry phase, shedding light on the topological phase \cite{Shoenberg,transport_review1,transport_review2,transport_review3}. However, determining the bulk band alignment is not easy in magneto-transport. In the vicinity of the NI to TI phase transition, the low-energy electronic structure of the material can be described to the lowest order of the wave vector $k$ as \cite{Murakami_1,Murakami_2}
\begin{align}
\label{Dirac}
E(k)=\alpha \sqrt{\hbar^2v_{F}^2k^2+M^2},
\end{align}
where $\alpha=\pm 1$ is the band index, $\hbar$ is the reduced Planck's constant, $v_F$ is the Fermi velocity, $M$ is the Dirac mass, and inversion symmetry is assumed. Although the amplitude of the bandgap can be measured as $|\Delta|=2M$, its sign ($+$ for normal and $-$ for inverted) cannot be determined using Eq. (1), as their corresponding band structures are identical.

In fact, band alignment makes differences only if the second order $k$ term ($\propto k^2$) is taken into account. In this work, we argue that when the contribution of the $k^2$ term is comparable to or larger than that of the Fermi velocity ($\propto k$), the band inversion manifests itself as a second energy extremum (bandgap) in the Brillouin zone, while the normal band remains a single extremum at the zone center. More saliently, we show that the second energy extremum in the band inversion case leads to a four-fold splitting of the density of states (DOS) in a magnetic field as well as an anomalous peak feature beyond the quantum limit. We experimentally confirm our model predictions using magneto-transport (including magneto-thermopower) measurements on zirconium pentatelluride (ZrTe$_5$), which is expected to be a strong TI (STI) with an inverted band at low temperatures. Our results not only supplement the popular magneto-transport technique with the decisive power in determining the band topology of the emergent topological materials, but also provide new perspectives in understanding their exotic behaviors.

\section*{Results}

\subsection*{Theoretical model and implications}
We start with the following model Hamiltonian for TIs \cite{BHZ_Zhang1,BHZ_Zhang2,BHZ_A3Bi,BHZ_Cd3As2,YJ_ZT5_PRB,BHZ_PBS,BHZ_ZT5}
\begin{align}
\label{Hamil}
H(\mathbf{k})= \hbar(v_{Fx}k_x\tau^x\sigma^z+v_{Fy}k_y\tau^y+v_{Fz}k_z \tau^x \sigma^x)+(M-\Sigma_{i}\mathscr{B}_i k^2_i) \tau^z,
\end{align}
where the Dirac mass $M$ of Eq. (1) is replaced by $M-\Sigma_{i}\mathscr{B}_i k^2_i$, $\mathscr{B}$ is the band inversion parameter, subscript $i$ denotes the crystal momentum direction, $\tau$ and $\sigma$ are the Pauli matrices for orbitals and spins, respectively. In typical narrow-gap materials, $\mathscr{B}\neq 0$ \cite{BHZ_Zhang1,BHZ_Zhang2}. For simplicity, we neglect the electron-hole asymmetry in our calculation and  also fix the sign for $M$ to be positive. Consequently, $\mathscr{B}>0$ represents the inverted band while $\mathscr{B}<0$ for the normal band.

At zero magnetic field, the energy dispersion along a particular $\mathbf{k}$ direction is then given by 
\begin{align}
\label{ZF}
E(k)=\alpha \sqrt{\hbar^2 v_{F}^2 k^2+(M- \mathscr{B} k^2)^2},
\end{align}
with $\mathbf{k}=0$ being the $\Gamma$ point. Taking $\mathbf{k}$ along the $z$ direction as an example, we plot the $k_z$ dispersion of the normal ($\mathscr{B}<0$, red) and inverted ($\mathscr{B}>0$, blue) bands in Fig. 1(a). We note that although only one energy extremum occurs at the $\Gamma$ point in the normal band case, a second extremum may appear at the $\zeta$ point for the inverted band. This can be better seen if we rewrite Eq. (3) as
\begin{align}
E(k_z)=\alpha \sqrt{\mathscr{B}_z^2(k_z^2-k_{\zeta}^2)^2+M^2-\mathscr{B}_z^2k_{\zeta}^4},
\end{align}
with $k_{\zeta}^2=(2M\mathscr{B}_z-\hbar^2v_{Fz}^2)/2\mathscr{B}_z^2$. For a second energy extremum at $k_z=k_{\zeta}$ to emerge, we require the band parameters to satisfy
\begin{align}
2M\mathscr{B}_z>\hbar^2v_{Fz}^2.
\end{align}
The bandgap at $k_z=k_{\zeta}$ is reduced to $|\Delta_{\zeta}|=2\sqrt{M^2-\mathscr{B}_z^2k_{\zeta}^4}<2M$ as compared to that of the $\Gamma$ point. Since condition (5) is only possible when the band is inverted ($\mathscr{B}_z>0$), the existence of a second extremum can serve as a criterion to identify band inversion \cite{YJ_ZT5_PRL}. We further note that even though our model can be broadly applied to most topological materials \cite{BHZ_Zhang1,BHZ_Zhang2,BHZ_Cd3As2,BHZ_A3Bi,BHZ_PBS,BHZ_ZT5} and broader topological phases \cite{PRB_Wang}, it is generally easier to satisfy the requirement for a discernible bandgap at the $\zeta$ point in layered materials such as ZrTe$_5$, HfTe$_5$, Bi$_2$Te$_3$, and Sb$_2$Te$_3$, as the weak coupling between layers naturally leads to a smaller $v_F$ along the layer stacking direction \cite{BHZ_Zhang2,YJ_ZT5_PRL,Thoery1_FZ}.

The presence of a second energy extremum at the $\zeta$ point has recently been confirmed using infrared (IR) spectroscopy \cite{YJ_ZT5_PRL,BSTS_WHH}. But, to the best of our knowledge, no transport evidence has yet been reported, mostly due to the difficulty in tuning the Fermi level in bulk crystals. On the other hand, when the system is near the NI to TI phase transition, because of the small bandgap, the two energy extrema at the $\Gamma$ and $\zeta$ points are very close to each other both in energy and $k_z$ separation, rendering their identification difficult at zero field even with high-resolution spectroscopic techniques. Nevertheless, such circumstances can be circumvented by the use of magnetic fields. In the following, we calculate the Landau level (LL) spectrum and DOS to search for the magneto-transport evidence for the inverted TI band. We restrict our discussion to the electron LLs, as our model preserves the electron-hole symmetry.

Figure 1(b) shows the $k_z$ dispersion of the low-lying LLs of the normal and inverted bands in Fig. 1(a) at $B=10$ T and applied along the $z$ direction. In the presence of a magnetic field, the electron states quantize into LLs, and the number of the energy extrema along the $k_z$ direction remains the same as zero field, that is, two for the inverted band at the $\Gamma$ and $\zeta$ points \cite{Ali_Cd3As22,PRL_HZL} but only one for the normal band at the $\Gamma$ point. When considering the $\mathscr{B} k^2$ term, the $n>0$ LLs split into two sub-levels even without the introduction of the Zeeman effect, and the splitting energy increases with larger $|\mathscr{B}|$. The Zeeman effect can change the energy separation of the two sub-LLs, but no additional splittings are expected (discussed in the Supplementary Note 2). Since each energy extremum produces a divergent contribution in DOS, calculated at the Fermi energy as a function of magnetic field (Fig. 1(c)), a total of four DOS peaks appear within each $n>0$ LL for the inverted band while only two peaks for the normal band. The $n=0$ LL, however, needs a separate discussion. First, it does not split under a magnetic field. Second, for a finite number of electrons, the Fermi level is always above the global minimum of the $n=0$ LL regardless of the magnitude of $B$, which is at the $\Gamma$ point for the normal band and the $\zeta$ point for the inverted band. Therefore, no DOS peak is expected in the $n=0$ LL of the normal band, whereas one peak for the inverted band (Fig. 1(c)) when the Fermi level is aligned with the $\Gamma$ point extremum. Details of the calculation can be found in Supplementary Notes 2.

The clear differences in the DOS between the normal and inverted bands can be well resolved in magneto-transport measurements. For example, the Shubnikov–de Haas oscillations (SdHOs) in magneto-resistance can be used to track the DOS oscillations of Fig. 1(c), where resistance/DOS peak occurs when the Fermi level sweeps through an energy extremum in Fig. 1(b) as varying the magnetic field. The concurrent presence of the four-fold splitting in the $n>0$ LLs and the anomalous peak in the $n=0$ LL can be viewed as direct evidence for the band inversion and the associated TI phase.

To test our model predictions, we choose as a case in point a popular topological material ZrTe$_5$, which exhibits a rich variety of novel properties, ranging from chiral magnetic effect \cite{Arpes0_GDG}, anomalous Hall effect \cite{Ong}, spin-zero effect \cite{JY_PNAS}, three-dimensional quantum Hall effect \cite{3DQH}, to saturating thermoelectric Hall effect \cite{PRL_MLT,NC_Zhang}. From the outset, theory predicts that ZrTe$_5$ is an STI but near the topological phase boundary to a weak topological insulator (WTI) \cite{Thoery1_FZ,NewTheory_Zhou}. Since WTI has a normal bulk bandgap \cite{Murakami_1,Murakami_2}, in this work we treat it interchangeably with NI. However, later studies reveal that depending on the growth method, the resulting crystals could be in different phases \cite{CJG}. Indeed, recent strain-dependent transport \cite{Strain_ZT5} and IR spectroscopy \cite{YJ_ZT5_PRL,PNAS_Chen,PRL_Xu} measurements of ZrTe$_5$ have both identified it as an STI at low temperatures. The magneto-IR measurement \cite{YJ_ZT5_PRL} has also demonstrated the presence of two energy gaps $\Delta_{\Gamma}=15\mbox{--}20$ meV and $\Delta_{\zeta}=9\mbox{--}12$ meV, separated by $k_{\zeta}\approx 0.17$ nm$^{-1}$, enabling the proposed test.

\subsection*{Experimental results}
Figure 2(a) shows the magneto-resistance of ZrTe$_5$, $\text{MR}\equiv [R_{xx}(B)/R_{xx}(0\text{T})-1]\times 100\%$, measured at different temperatures. At low field, the MR grows almost linearly with $B$ up to 10 T, and it continues to grow between 10-31 T, particularly at low temperatures. At the lowest temperature of $T=2.3$ K, $\text{MR}=3900\%$ at 31 T. In addition to the increasing MR, SdHOs can be recognized in Fig. 2(a). As the temperature increases, the oscillation vanishes at 50 K. Therefore, we can extract the oscillatory component $\Delta R_{xx}(2.3\text{K})$ by subtracting a smooth background using the $R_{xx}(50\text{K})$ curve (Fig. 2(b)) \cite{PRB_Nair}. In Fig. 2(c), we replot $\Delta R_{xx}(2.3\text{K})$ as a function of $1/B$. By extracting the magnetic field positions of the peaks and valleys in SdHOs, we construct a Landau fan diagram in Fig. 2(d) and determine the corresponding LL index $n$. From Fig. 2(d), we can also deduce the frequency of SdHOs, $F=3.6$ T, consistent with previous results \cite{JY_PNAS,SR_Yu,PRB_TML,g_FXX,g_XHC}.

The most striking feature of SdHOs is the prominent peak, labeled by $n=0$ in Figs. 2(b) and 2(c), centered at 23 T. From the Landau fan diagram (Fig. 2(d)), we can identify that the system enters the quantum limit ($n=1$ LL) around 10 T. An additional peak in $\Delta R_{xx}$ in the extreme quantum limit is thus not expected for the normal band. Instead, as we have discussed above, the emergence of the $n=0$ peak can be attributed to the presence of two energy extrema in the inverted band (Fig. 1). This interpretation is further supported by the observation of a board $n=1$ peak in Figs. 2(b) and 2(c) and a possible LL splitting (labeled by the $\star$ symbol) around 5 T \cite{SR_Yu}. The splitting behavior is more pronounced when we obtain $\Delta R_{xx}$ by subtracting a non-oscillating polynomial background, as shown in Supplementary Figure S2. In Fig. 2(d), the $n=1$ peak data also appear deviated from the fitting using the peaks and valleys of higher LL indices. This deviation can be corrected if we consider a (four-fold) splitting within the $n=1$ LL, which will be further discussed later.

Fortunately, we can better resolve the $n=1$ peak via magneto-thermopower measurements. It is well known that thermopower is proportional to the derivative of the conductivity to energy (Mott relation) \cite{Fu_1,Fu_2}. Therefore, it is more sensitive to fine structures in the DOS, such as splittings. Figure 3(a) shows the normalized magneto-thermopower, $S_{xx}(B)/S_{xx}(0\text{T})$, measured at selected temperatures. At the lowest temperature of 2.3 K, $S_{xx}$ exhibits quantum oscillations in low magnetic fields, reaching the quantum limit ($n=1$ LL) around 7.4 T, consistent with the $R_{xx}$ measurements above. In Fig. 3(b), we plot the oscillatory component $\Delta S_{xx}$ as a function of $1/B$ and observe clear splitting behavior, as indicated by the down-triangles. Particularly, for the $n=1$ LL, three marked peaks are evidenced. We reiterate that such splitting cannot be explained by the Zeeman effect in the normal band picture, as it only changes the energy separation between the upper and lower sub-LLs, but the splitting remains as two-fold (discussed in Supplementary Note 2). We, therefore, attribute it to the four-fold splitting in the inverted band, with the additional two-fold resulting from the two energy extrema at the $\Gamma$ and $\zeta$ points.

Figure 3(c) shows the Landau fan diagram, in which we group the split peaks into two branches labeled by the red and blue down-triangles. To identify the four-fold splitting within the $n=1$ LL, we linearly fit the splitting in higher ($n>1$) LLs and deduce the dominant splitting at $n=1$. The dash lines in Fig. 3(c) thus represent the average positions of two blue and red down-triangles in Fig. 3(b), that is, the average positions of additional splittings in the $n=1$ LL. The inner two peaks of the four-fold splitting are at similar magnetic fields, therefore merged into one in the experimental data. Finally, the slope of the blue and red lines in the Landau fan diagram of Fig. 3(c) are 3.8 T and 4.2 T, respectively, giving rise to an average value of 4 T, consistent with the MR measurements.

At 8 T, we observe a dramatic drop in $S_{xx}$ that can be attributed to the drop in DOS when the Fermi level moves away from the bottom of the $n=1$ LL \cite{PRL_MLT,Drop1,Drop2}. As we further increase the magnetic field, the $S_{xx}$ value remains low and forms a plateau region between $15<B<19$ T, before it grows again forming the $n=0$ hump in the extreme quantum limit. We note that the observed plateau region is consistent with that reported in the previous high-field magneto-thermopower measurements on ZrTe$_5$ \cite{PRL_MLT}, and from the calculated DOS in Fig. 1(c), it corresponds to the DOS minimum before entering the $n=0$ LL orbit (see also the calculated $S_{xx}$ in Fig. 4(a)).

The hump/peak feature of the $n=0$ LL can be better seen if we normalize $S_{xx}(2.3\text{K})$ to $S_{xx}(50\text{K})$, as shown in the inset to Fig. 3(b). After the $n=0$ hump, we also observe a slightly negative $S_{xx}$ around 27 T. We suspect that the application of a high magnetic field induces hole carriers in ZrTe$_5$ at the Fermi surface, which is hinted by a concurrent sign reversal in the slope of the Hall data at high magnetic fields, as shown in Supplementary Figure S3(a).

\section*{Discussion}
There are different ways to assign three peaks to a four-fold splitting, depending on the sequence in which the Fermi level sweeps through the band extrema. In Fig. 3(b), we use red and blue down-triangles to illustrate two possible assignments: (1) Red and blue represent the splitting of the upper and lower sub-LLs ($\Delta_s$). In this case, the sub-LL splitting is larger than the $\Gamma$ and $\zeta$ point splitting in energy, $\Delta_s>\Delta_{\Gamma\zeta}$ (as shown in the inset to Fig. 4(a)). Therefore, the Fermi level first touches the two extrema in the upper sub-LLs and then the lower ones as the magnetic field increases. (2) Alternatively, we can attribute the red and blue down-triangles to the $\Gamma$ and $\zeta$ point energy splitting, if it is larger than the sub-LL splitting, $\Delta_{\Gamma\zeta}>\Delta_s$. Here, as increasing the magnetic field, the Fermi level first sweeps through the $\Gamma$ point energy extremum in each sub-LL and then the two $\zeta$ point extrema. Further discussion of the LL hierarchy can be found in Supplementary Note 3.

A closer inspection of the splitting behavior in Fig. 3(c) tentatively suggests that assignment (1) may be the possible scenario. In this case, since the dominant sub-LL splitting is due to a Zeeman-like behavior, we can linearly fit the quantum oscillations in the upper (red down-triangles) and lower (blue down-triangles) sub-LLs and extract an effective $g$-factor. Specifically, we can describe the splitting as $g_{eff}m^*/2m_e=F\Delta(1/B)$, where $m^*$ ($m_e$) is the effective (bare) electron mass, and $\Delta(1/B)$ is the spacing between the split peaks. In our ZrTe$_5$ samples, $F=4.0$ T from the fitting in Fig. 3(c) and $m^*=0.045m_e$ from the temperature-dependent MR measurements (detailed in Supplementary Note 1). The deduced electron $g$-factor is then about $g_{eff}\approx 11$ using the splitting of the $n=2$ peak, which is smaller than the previous studies \cite{g_FXX,g_XHC} and may suggest sample dependence. On the other hand, in assignment (2), the red and blue down-triangles are attributed to the $\Gamma$ and $\zeta$ point energy splitting, which is strongly dependent on the band parameter differences at these two points. The slopes of the two linear fits in Fig. 3(c) are thus expected to be proportional to the corresponding Fermi surface area, which is approximately $\propto 1/v_F^2$ at the $\Gamma$ and $\zeta$ points. However, in Fig. 3(c), we find the ratio of the two slopes (red/blue) $\approx 1.1>1$, whereas our previous magneto-IR spectroscopy measurements reveal a ratio of $v^2_{F,\zeta}/v^2_{F,\Gamma}\approx 0.82<1$ \cite{YJ_ZT5_PRL}. 

Lastly, we show in Fig. 4, the numerically calculated $R_{xx}$ and $S_{xx}$ using our theoretical model and a Green's function method (detailed in Supplementary Note 4). By assuming a constant scattering rate of 0.6 meV and a weakly varied $k$-dependent Fermi velocity \cite{YJ_ZT5_PRL}, a qualitative agreement between the calculation and experiment is achieved. Specifically, we find that for $n>1$ LLs, the energy broadening only allows for the identification of a doublet structure in $R_{xx}$ and $S_{xx}$, as their sub-LL splitting in field is relatively small, while a four-fold splitting occurs in the $n=1$ LL, as indicated by the down-triangles in Fig. 4(a). After the $n=1$ LL, a sharp drop occurs in both $R_{xx}$ and $S_{xx}$, followed by a plateau region and the emergence of the $n=0$ anomalous peak in the extreme quantum limit. The continued increase in $R_{xx}$ after the $n=0$ peak may have been subtracted from the background in the experimental data of Figs. 2(b) and 2(c). In this calculation, we incorporate the Zeeman effect and a total effective $g$-factor of $g_{eff}\approx 15$, slightly larger than the value $g_{eff}\approx 11$ deduced above from the experiment. Overall, the calculation is in good qualitative agreement with the experiment. A more quantitative description of the experiment, such as the width of the plateau region and the magnetic field, and the energy-dependent broadening that could lead to the triplet structure in the $n=1$ peak, would require accurate $k$-dependent band parameters and information on the scattering mechanism, therefore beyond the scope of this work. More calculation details can be found in Supplementary Note 4.

In conclusion, we have demonstrated both theoretically and experimentally that the band inversion in anisotropic topological material ZrTe$_5$ can lead to a second energy extremum in the electronic structure and result in unique electronic responses. These include a four-fold splitting of the $n=1$ peak in the magneto-thermopower quantum oscillations as well as an anomalous $n=0$ peak in the extreme quantum limit. The concurrent presence of these characteristic features can rigorously distinguish the inverted band structure from the normal band in narrow-gap materials, further promoting magneto-transport measurement, particularly high-field magneto-thermopower measurement, as a powerful tool in studying the band topology.\\

\section*{Methods}
The ZrTe$_5$ samples were grown by the chemical vapor transport method \cite{SR_Yu}. ZrTe$_5$ polycrystals were first prepared by reacting the appropriate ratio of Zr and Te in a vacuumed quartz tube at 450 $^\circ$C for one week. Then, ZrTe$_5$ single crystals were grown using iodine as the transport agent, and the transport temperature was from 530 $^\circ$C to 450 $^\circ$C. The transport time was around 20 days.

For transport measurements, we mechanically exfoliated the sample down to a 188-nm-thin flake and deposited it onto the Si/SiO$_2$ substrate. Due to its layered, anisotropic structure, the ZrTe$_5$ thin flake is in the $ac$ plane with the long direction being along the $a$ axis. We then patterned Pd/Au (5/150 nm thick) contacts in Hall-bar geometry using standard electron-beam lithography. To achieve ohmic contacts, we employed Ar plasma etching before the metal deposition. An optical microscope image of our device is shown in the inset to Fig. 2(a). The electronic transport measurements are carried out by standard lock-in method  with excitation current frequency $f_R=17.777$ Hz in a variable temperature cryostat equipped with a resistive magnetic ($B\parallel b$ axis) up to 31 T.

For thermopower measurements, a micro-heater (Ti/Au, 5/45 nm thick) was also fabricated to provide a longitudinal thermal gradient along the crystal $a$ axis. To obtain a high signal-to-noise ratio, we employ a second harmonic method \cite{ZJ_thesis}. The frequency of the alternating current through the micro-heater is $f_H=3.777$ Hz. Consequently, the frequency of the induced temperature oscillation is $2f_H$ with a phase shift of $-\pi/2$. $S$ is proportional to the second harmonic component of the thermoelectric voltage at phase $-\pi/2$.

\section*{Data Availability}
The data that support the findings of this study are available from the corresponding author upon reasonable request.\\

\section*{Supplementary Information}
\noindent Supplementary information for this article is available at Link insert by editorial team.\\

\section*{Acknowledgments}
This work was primarily supported by the DOE through Grant No. DE-FG02-07ER46451, while crystal growth at GT and UT were supported by Grant No. DE-SC0018660 and DE-SC0020254. Crystal characterization was performed in part at the GT Institute for Electronics and Nanotechnology, a member of the National Nanotechnology Coordinated Infrastructure, which is supported by the NSF (Grant No. ECCS-2025462). High-field transport measurements were performed at the NHMFL, which is supported by the NSF Cooperative Agreement No. DMR-1644779 and the State of Florida. J.W. acknowledges support from the China Scholarship Council. Y.J. acknowledges support from the NHMFL Jack Crow Postdoctoral Fellowship, and Z.J. acknowledges the NHMFL Visiting Scientist Program. A.L.M. and Z.J. acknowledge support from the Sandia Academic Alliance Program. Work at Sandia is supported by a Laboratory Directed Research and Development project and a user project at the Center for Integrated Nanotechnologies, an Office of Science User Facility operated for the DOE Office of Science. Sandia National Laboratories is a multimission laboratory managed and operated by National Technology and Engineering Solutions of Sandia, LLC., a wholly owned subsidiary of Honeywell International, Inc., for the U.S. DOE's National Nuclear Security Administration under contract DE-NA-0003525. This paper describes objective technical results and analysis. Any subjective views or opinions that might be expressed in the paper do not necessarily represent the views of the U.S. DOE or the U.S. Government.\\

\section*{Author contributions}
Y.J. and Z.J. conceived the projects. Z.D., M.M., and H.Z. grew the ZrTe$_5$ crystals. J.W., T.Z., and A.L.M. fabricated the devices. J.W., T.Z., and Y.J. performed the experiments. Y.J. performed the calculations. X.W., W.P., D.S., and Z.J. supervised the project. J.W., Y.J., and Z.J. wrote the manuscript and all authors contributed to the manuscript.\\

\section*{Competing interests:}
The authors declare that they have no competing financial interests.\\

\clearpage

\begin{figure}[t!]
\centering
\includegraphics[width=9.5cm]{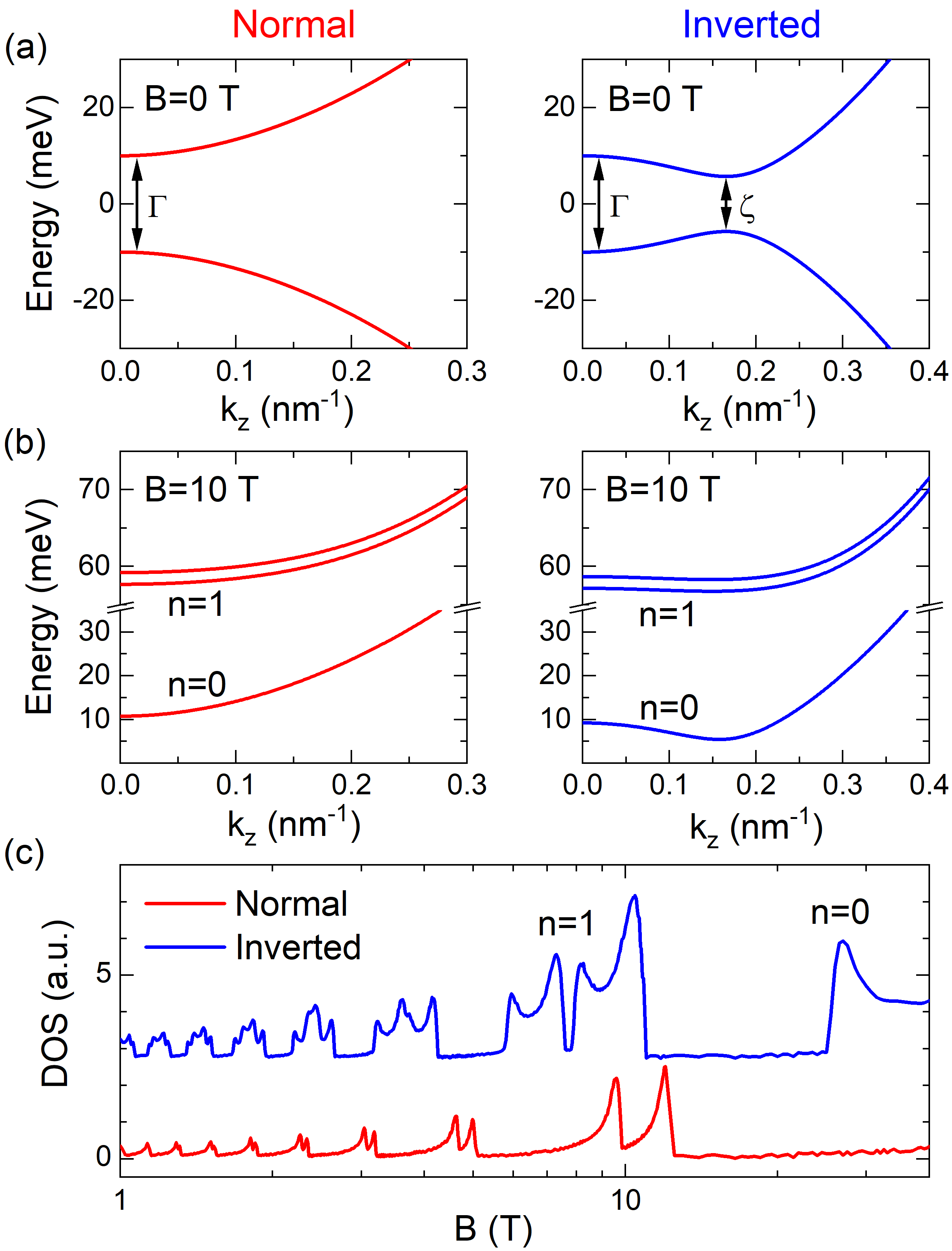}
\end{figure}
\noindent Fig. 1. Second bandgap induced by band inversion. (a) Side-by-side comparison of the zero-field $k_z$ dispersion of the normal (red) and inverted (blue) bands. The dispersions are produced using Eq. \eqref{ZF} with practical band parameters \cite{YJ_ZT5_PRL}, $v_{Fz}=5 \times 10^4 $ m/s, $M=10$ meV, $\mathscr{B}_z=-0.3$ eV$\ $nm$^2$ (normal), and $\mathscr{B}_z=0.3$ eV$\ $nm$^2$ (inverted), leading to a second energy extremum at the $\zeta$ point for the inverted band. (b) $k_z$ dispersion of the low-lying electron LLs of the normal and inverted bands in (a) at $B=10$ T. Here, $\mathbf{B}\parallel z$ direction, and we use an average value for the band parameters in the $x$-$y$ plane, including $v_{F,\perp}=5\times 10^5$ m/s and $\mathscr{B}_{\perp}=0.05$ eV$\ $nm$^2$. Integer $n$ denotes the LL index. (c) Calculated DOS oscillations as a function of magnetic field with a constant electron density of $4 \times 10^{17}$ cm$^{-3}$.
\clearpage

\begin{figure}[t!]
\centering
\includegraphics[width=12cm]{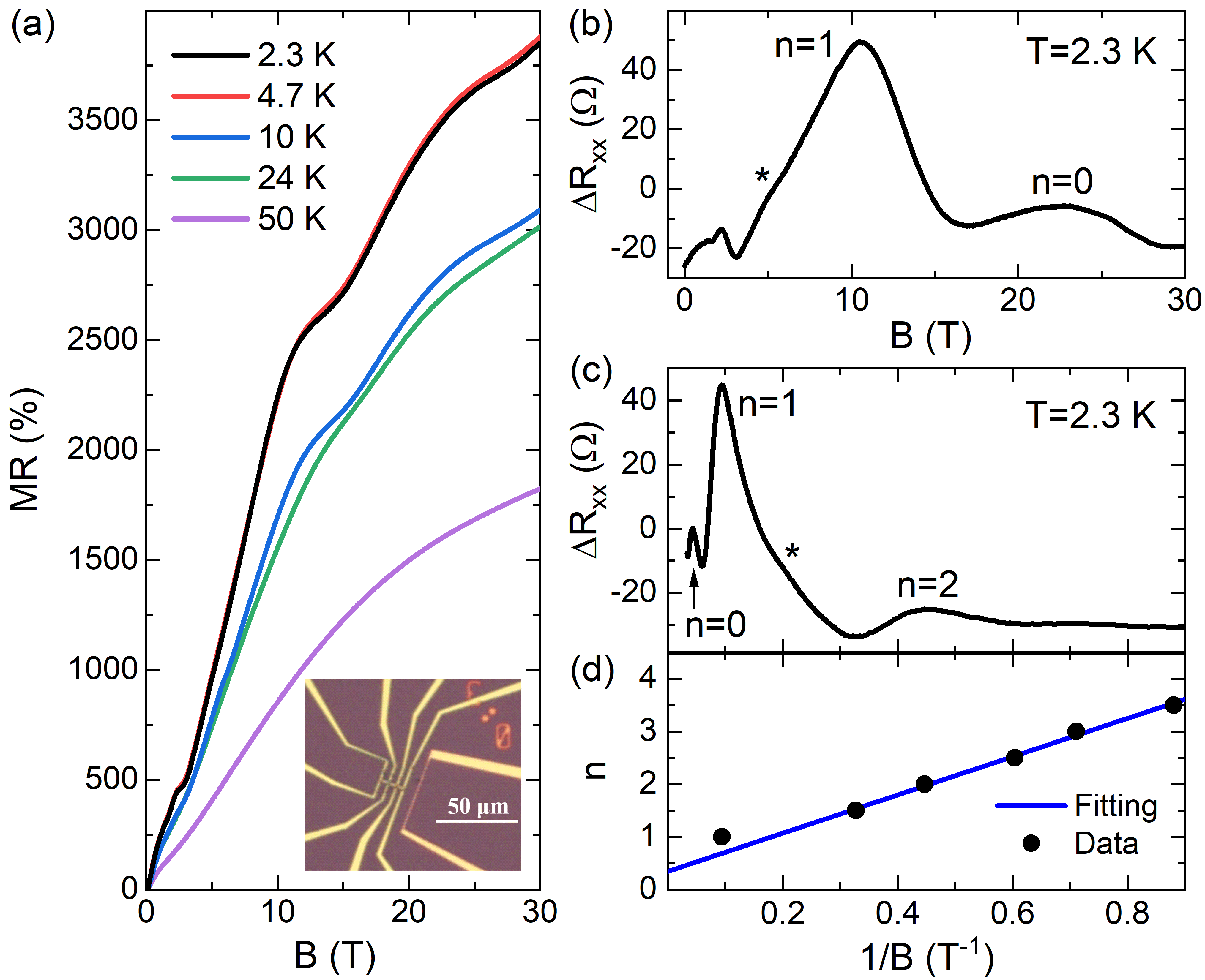}
\end{figure}
\noindent Fig. 2. Magneto-resistance measurements in ZrTe$_5$. {(a) Magneto-resistance, $\text{MR}\equiv$\\$[R_{xx}(B)/R_{xx}(0\text{T})-1]\times 100\%$, measured at different temperatures. Inset: Optical microscope image of a device designed for both MR and magneto-thermopower measurements. (b) Quantum oscillations in $\Delta R_{xx}$ as a function of $B$ at 2.3 K. (c) $\Delta R_{xx}$ as a function of $1/B$ at 2.3 K. $\Delta R_{xx}$ is obtained by subtracting a smooth background (we choose $R_{xx}(50\text{K})$ as the background) from $R_{xx}$, $\Delta R_{xx}=R_{xx}(2.3\text{K})-R_{xx}(50\text{K})$. The $\star$ symbol indicates a possible splitting of the $n=1$ peak. (d) Landau fan diagram of the SdHOs, with the peaks (valleys) in $\Delta R_{xx}$ assigned to $n$ ($n+1/2$), respectively. The straight line is a linear fit to the data at $n>1$.}
\clearpage

\begin{figure}[t!]
\centering
\includegraphics[width=12cm]{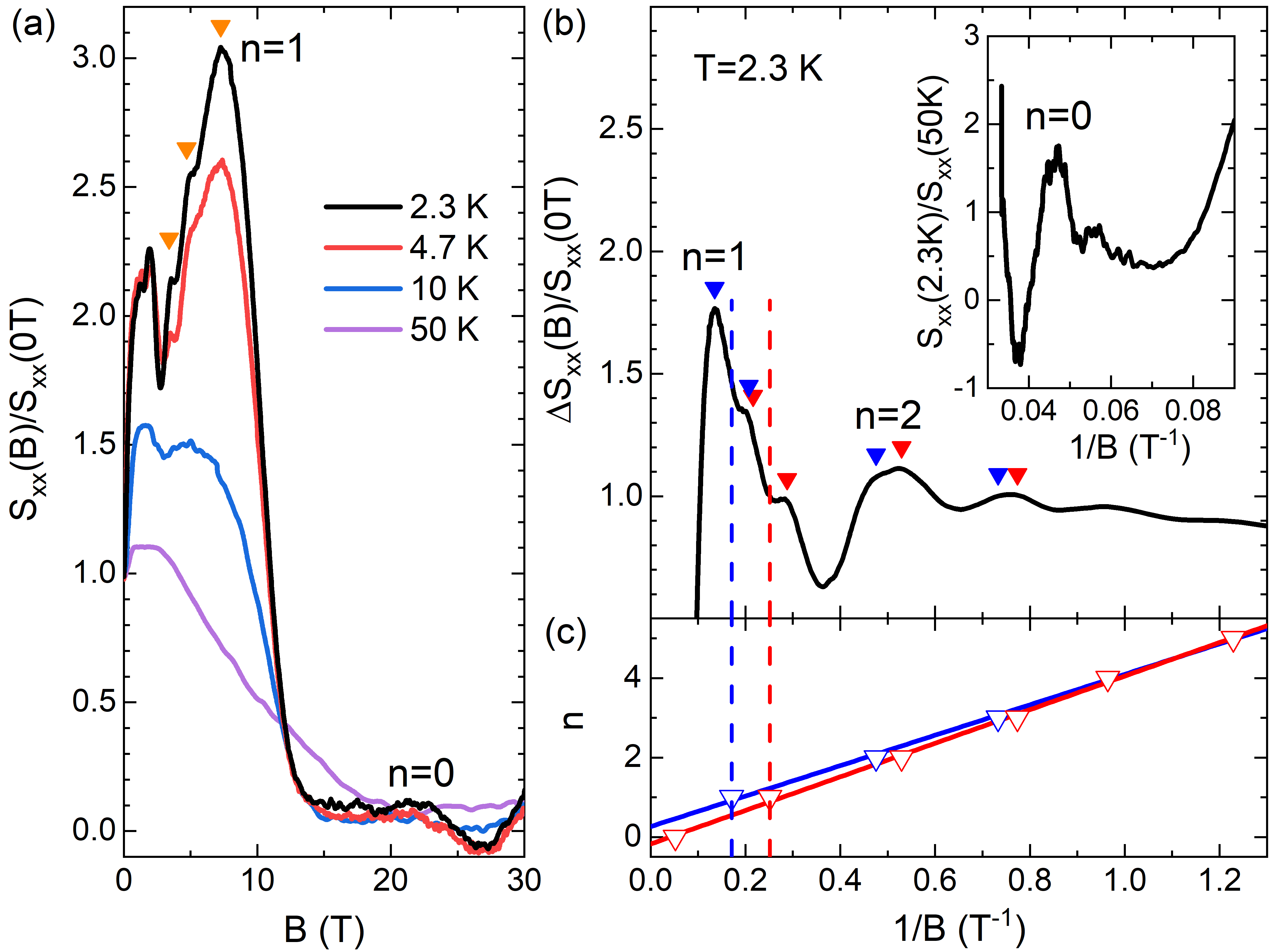}
\end{figure}
\noindent Fig. 3. Magneto-thermopower measurements in ZrTe$_5$.{ (a) Normalized magneto-thermopower, $S_{xx}(B)/S_{xx}(0\text{T})$, measured at different temperatures. The three orange down-triangles indicate the splitting of the $n=1$ peak. (b) Oscillatory component $\Delta S_{xx}$ as a function of $1/B$ at 2.3 K, obtained by subtracting a linear background from $S_{xx}$. Inset: The $n=0$ peak obtained by reference to the high-temperature data at 50 K, $S_{xx}(2.3\text{K})/S_{xx}(50\text{K})$. (c) Landau fan diagram of the magneto-thermopower oscillations at 2.3 K. The peak features in (b) are separated into two groups depending on the splitting mechanism and labeled by the red and blue down-triangles, respectively. The dash lines indicate the average positions of additional splittings in the $n=1$ LL, and the solid lines are linear fits to the data of each group.}
\clearpage

\begin{figure}[t!]
\centering
\includegraphics[width=12cm]{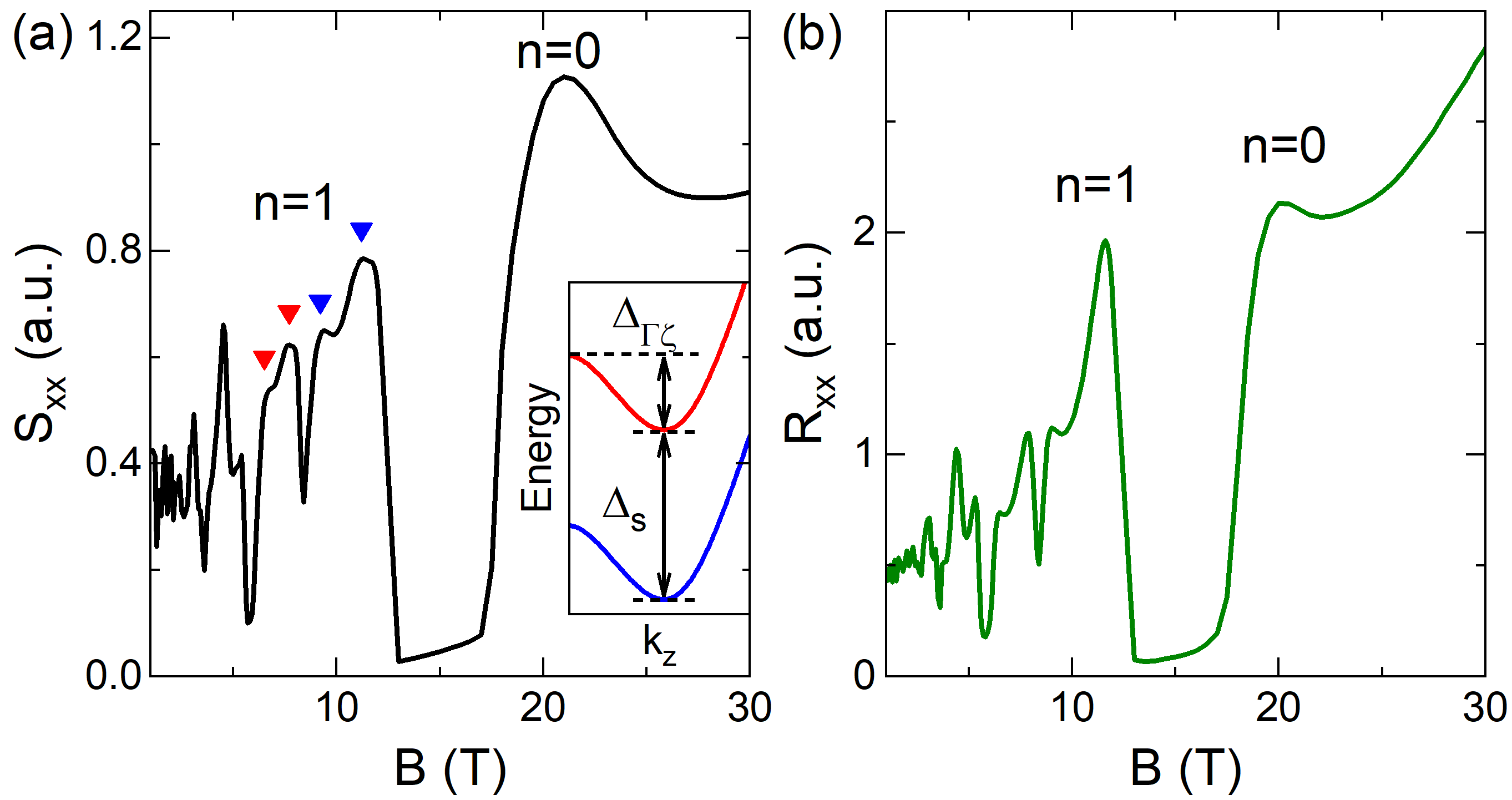}
\end{figure}
\noindent Fig. 4. Calculated magneto-resistance and magneto-thermopower. {Calculated (a) $S_{xx}$ and (b) $R_{xx}$ as a function of magnetic field. To demonstrate the four-fold splitting, a relatively small scattering rate of 0.6 meV is used in the calculation. Other parameters are listed in Supplementary Note 4. The red and blue down-triangles mark the four-fold splitting of the $n=1$ peak in $S_{xx}$, following the assignment in Fig. 3(b). Inset to (a) illustrates the band alignment within the $n=1$ LL, according to assignment (1) described in the main text.}
\end{document}